\documentclass[manuscript]{aastex}

\shorttitle{Formation conditions of comet 67P/Churyumov-Gerasimenko}
\shortauthors{Mousis et al.}

\usepackage{etex}
\usepackage{m-pictex, m-ch-en}
\usepackage {amsmath}
\usepackage[squaren, Gray, cdot]{SIunits}
\usepackage{color}

\begin{document}


\title{Impact of radiogenic heating on the formation conditions of comet 67P/Churyumov-Gerasimenko}


\author{O. Mousis\altaffilmark{1}, A. Drouard\altaffilmark{1}, P. Vernazza\altaffilmark{1}, J. I. Lunine\altaffilmark{2}, M., Monnereau\altaffilmark{3}, R. Maggiolo\altaffilmark{4}, K. Altwegg\altaffilmark{5,6}, H. Balsiger\altaffilmark{5}, J.-J. Berthelier\altaffilmark{7}, G. Cessateur\altaffilmark{4}, J. De Keyser\altaffilmark{4}, S. A. Fuselier\altaffilmark{8}, S. Gasc\altaffilmark{4}, A. Korth\altaffilmark{9}, T. Le Deun\altaffilmark{1}, U. Mall\altaffilmark{9}, B. Marty\altaffilmark{10}, H. R\`eme\altaffilmark{3}, M. Rubin\altaffilmark{5}, C.-Y. Tzou\altaffilmark{5}, J. H. Waite\altaffilmark{8}, and P. Wurz\altaffilmark{5}}


\altaffiltext{1}{Aix Marseille Universit{\'e}, CNRS, LAM (Laboratoire d'Astrophysique de Marseille) UMR 7326, 13388, Marseille, France {\tt olivier.mousis@lam.fr}}
\altaffiltext{2}{Department of Astronomy and Carl Sagan Institute, Space Sciences Building Cornell University,  Ithaca, NY 14853, USA}
\altaffiltext{3}{Universit\'e de Toulouse; UPS-OMP-CNRS; IRAP, Toulouse, France}
\altaffiltext{4}{Royal Belgian Institute for Space Aeronomy, BIRA-IASB, Ringlaan 3, B-1180 Brussels, Belgium}
\altaffiltext{5}{Physikalisches Institut, University of Bern, Sidlerstr. 5, CH-3012 Bern, Switzerland}
\altaffiltext{6}{Center for Space and Habitability, University of Bern, Sidlerstr. 5, CH-3012 Bern, Switzerland}
\altaffiltext{7}{LATMOS/IPSL-CNRS-UPMC-UVSQ, 4 Avenue de Neptune F-94100, Saint-Maur, France}
\altaffiltext{8}{Department of Space Science, Southwest Research Institute, 6220 Culebra Rd., San Antonio, TX 78228, USA}
\altaffiltext{9}{Max-Planck-Institut f\"ur Sonnensystemforschung, Justus-von-Liebig-Weg 3, 37077 G\"ottingen, Germany}
\altaffiltext{10}{Centre de Recherches P\'etrographiques et G\'eochimiques, CRPG-CNRS, Universit\'e de Lorraine, 15 rue Notre Dame des Pauvres, BP 20, 54501 Vandoeuvre l\`es Nancy, France}

\begin{abstract}
Because of the high fraction of refractory material present in comets, the heat produced by the radiogenic decay of elements such as aluminium and iron can be high enough to induce the loss of ultravolatile species such as nitrogen, argon or carbon monoxide during their accretion phase in the protosolar nebula. Here, we investigate how heat generated by the radioactive decay of $^{26}$Al and $^{60}$Fe influences the formation of comet 67P/Churyumov-Gerasimenko, as a function of its accretion time and size of parent body. We use an existing thermal evolution model that includes various phase transitions, heat transfer in the ice-dust matrix, and gas diffusion throughout the porous material, based on thermodynamic parameters derived from {\it Rosetta} observations. Two possibilities are considered: either, to account for its bilobate shape, 67P/Churyumov-Gerasimenko was assembled from two primordial $\sim$2 kilometer-sized planetesimals, or it resulted from the disruption of a larger parent body with a size corresponding to that of comet Hale-Bopp ($\sim$70 km). To fully preserve its volatile content, we find that either 67P/Churyumov-Gerasimenko's formation was delayed between $\sim$2.2 and 7.7 Myr after that of Ca-Al-rich Inclusions in the protosolar nebula or the comet's accretion phase took place over the entire time interval, depending on the primordial size of its parent body and the composition of the icy material considered. Our calculations suggest that the formation of 67P/Churyumov-Gerasimenko is consistent with both its accretion from primordial building blocks formed in the nebula or from debris issued from the disruption of a Hale-Bopp-like body.

\end{abstract} 


\section{Introduction}
Radiogenic heating has played a major role in the evolution of small bodies in the early solar system (Shukolyukov \& Lugmair 2002; Formisano et al. 2013). Given the fact that short-lived nuclides such as $^{26}$Al and $^{60}$Fe were present in these bodies, they may have constituted a major heat source for metamorphism, melting, and differentiation in asteroids (Grimm \& McSween 1989; Ghosh \& McSween 1998; McSween et al. 2002; Huss 2004; Monnereau et al. 2013a). Radiogenic heating may have generated temperatures high enough in the interiors of Kuiper Belt Objects to crystallize amorphous ice, the melting of water ice and the loss of ultravolatiles (Choi et al. 2002; De Sanctis et al. 2001; Merk \& Prialnik 2003, 2006; Prialnik et al. 2008; Guilbert-Lepoutre et al. 2011). The influence of radiogenic heating has also been explored in comets. It was found that these bodies had to accrete over several Myr before reaching their final sizes to retain their amorphous ice, assuming they agglomerated from this solid phase (Prialnik et al. 1987; Prialnik \& Podolak 1995). Radiogenic heating could have even been at the origin of the oligomerization of molecules such as HCN and NH$_3$ to form amino acids (Yabushita 1993). The nitrogen deficiency observed in these bodies could result as well from the internal heating engendered by the decay of $^{26}$Al and $^{60}$Fe present in the refractory phase (Mousis et al. 2012). Meanwhile, formation delays of several Myr after the formation of Ca-Al-rich Inclusions (CAIs) in the protosolar nebula (PSN) have been invoked to maintain the presence of carbon monoxide in comets with sizes similar to that of Hale-Bopp (Mousis et al. 2012; Monnereau et al. 2013b).

Dynamical simulations representing the collisional evolution of planetesimal disks suggest that the Jupiter Family Comets (JFCs) are predominantly fragments resulting from collisions experienced by larger parent bodies (Davis \& Farinella 1997; Schlichting et al. 2013; Morbidelli \& Rickman 2015). On the other hand, based on differences observed between Trans-Neptunian Objects and comets (presence of aqueous alteration in larger bodies, differences in degrees of compaction, etc.), Davidsson et al. (2016) recently proposed that comet nuclei correspond to primordial rubble piles rather than being fragments of collisions.

Here, we investigate how heat generated by the radioactive decay of $^{26}$Al and $^{60}$Fe influences the formation of comet 67P/Churyumov-Gerasimenko (67P/C-G), as a function of its accretion delay (after CAI formation) and size of parent body. To do so, we use a thermal evolution model that includes various phase transitions, heat transfer in the ice-dust matrix, and gas diffusion throughout the porous material, based on Marboeuf et al. (2012) and on thermodynamic parameters derived from {\it Rosetta} observations. Two possibilities are considered: either, to account for its bilobate shape, 67P/C-G was assembled from two primordial $\sim$2 kilometer-sized planetesimals, or it results from the disruption of a larger parent body with a size corresponding to that of comet Hale-Bopp. 


\section{Thermal evolution of 67P/C-G}
\subsection{Nucleus model}

We use the one-dimensional thermal evolution model presented in Marboeuf et al. (2012), which has been previously utilized to depict the formation of pits on the surface of 67P/C-G (Mousis et al. 2015) and to characterize the subsurface of the ESA/{\it Rosetta} descent module {\it Philae} landing site (Brugger et al. 2016). In this model, the nucleus consists of a sphere made of a porous mixture of water ice and other volatile molecules (in both gas and solid states), along with dust grains in specified proportions. The model describes heat transfer, latent heat exchanges, all possible water ice structures (crystalline ice, amorphous ice, and clathrate) and phase changes (amorphous-to-crystalline ice transition, clathrate formation from crystalline or amorphous ice and crystalline ice formation from clathrate destabilization), sublimation/recondensation of volatiles in the nucleus, gas diffusion, gas trapping or release by clathrate formation or dissociation, as well as gas and dust release and mantle formation at the nucleus surface. 

The model computes the time evolution of the temperature distribution by solving the heat diffusion equation:

\begin{equation}
\rho c \frac{\partial T}{\partial t} = \boldsymbol{\nabla} \cdot \left(K^m \boldsymbol{\nabla} T\right) + Q,
\label{eq1}
\end{equation}

\noindent where {\it T} is the temperature (K), {\it t} the time (s), {\it r} the distance (m) from the center of the body, $\rho$ the mean density of the nucleus (kg~m$^{-3}$), {\it c} its mean specific heat (J~kg$^{-1}$~K$^{-1}$) and $K^m$ the heat conduction coefficient (J~s$^{-1}$~m$^{-1}$~K$^{-1}$) of the porous matrix. $Q$ corresponds to the amount of power per unit volume (J s$^{-1}$ m$^{-3}$) supplied to or released from the porous matrix. This term can be broken down as follows:

\begin{equation}
Q = Q^g + Q^{cr} + Q^{ga} + Q^{cl} + Q^{rad},
\label{eq2}
\end{equation}

\noindent with $Q^g$ the global power per unit volume resulting from the different phase changes experienced by molecules at the surface of the pores (condensation, adsorption or sublimation), $Q^{cr}$ the power per unit volume released during the crystallization of amorphous ice, $Q^{ga}$ the power per unit volume exchanged between the different molecules in the gas phase, which diffuse within the solid matrix via its porous network, $Q^{cl}$ the power per unit volume released/taken up during the formation/dissociation of clathrate cages, and $Q^{rad}$ the power per unit volume released by radiogenic heating within the matrix. Except for the term $Q^{rad}$, which has been added to the model and is described below, we refer the reader to Marboeuf et al. (2012) for a full description of the model.

\subsection{Radiogenic heating}

We assume the presence of the short-lived nuclides $^{26}$Al and $^{60}$Fe in the dust fraction considered in our model. The power per unit volume supplied to dust by radioactive decay can then be quantified via the following relation:

\begin{equation}
Q^{rad}_{dust} = \sum_{rad} \rho_d X_{rad}(t_D) H_{rad} \frac{1}{\tau_{rad}}~\exp\left(\frac{-t}{\tau_{rad}}\right),
\label{eq3}
\end{equation}

\noindent with $\rho_d$ the dust density (kg~m$^{-3}$), $X_{rad}$($t_D$) the mass fraction of a radioactive isotope in the dust assuming a given formation delay $t_D$ of the nucleus after CAI formation, $H_{rad}$ the heat released per unit mass (J~kg$^{-1}$) upon decay and $\tau_{rad}$ the mean lifetime (s). Decay during the formation delay $t_D$ results in a decrease of each nuclide's initial abundance:

\begin{equation}
X_{rad} (t_D) = X_{rad} (0)~\exp\left(\frac{-t_D}{\tau_{rad}}\right),
\label{eq4}
\end{equation}

\noindent with $X_{rad}(0)$ the initial mass fraction of each nuclide. $Q^{rad}$ is then derived from $Q^{rad}_{dust}$ as a function of porosity and dust-to-ice ratio. Our thermal evolution model does not account for the growth of the body during its accretion phase. Consequently, the nucleus accretion time is assumed to be small compared to $t_D$. The computation of the thermal evolution starts at time zero after the delay $t_D$, from an initial temperature of 30 K, which corresponds to the surface temperature of a planetesimal orbiting the Sun at a distance of $\sim$85 AU. No additional accretional heating is accounted for.

\subsection{Parameters}

Our computations have been conducted under the assumption that 67P/C-G results from the merging of two lobes originally formed separately (Davidsson et al. 2016). Therefore, two extreme body sizes have been considered. In the first case, we assumed that these lobes are primordial and reached their current sizes at the end of their accretion. We then used an average value inferred from the measured sizes of 67P/C-G's lobes. In the second case, we postulated that these lobes originated from the disruption of larger bodies (Morbidelli \& Rickman 2015). Consequently, we adopted a generic Hale-Bopp-like size for the body under consideration, a value close to the average sizes of P-- and D--types asteroids as well as of Jovian Trojans, which are good candidates for comets' parent bodies (Vernazza et al. 2015; Vernazza \& Beck 2017). Two distinct ice structures, based on the current literature (Bar-Nun et al. 2007; Luspay-Kuti et al. 2016; Mousis et al. 2016), have been investigated for each size:

\begin{itemize}

\item {\it Mixed model.} The icy phase is made of pure solid water distributed half as pure crystalline ice and half in clathrate form. Clathrate destabilization is simulated without any volatile inclusion in the cages; the latter weakly affects the energetics of the destabilization process.

\item {\it Amorphous model.} The icy phase of the nucleus is exclusively made of pure amorphous water ice.

\end{itemize}

Water is the only volatile species considered in our model, allowing the computational time of each simulation to be significantly reduced (days vs. weeks/months for some simulations). Finally, two values of dust-to-ice ratios, namely 4 and 1, have been investigated in our simulations. The higher and lower values correspond to those measured in 67P/C-G (Rotundi et al. 2015) and assumed in the case of a more primitive body, respectively. Table 1 displays the key parameters used in our simulations. Other structural and thermodynamic parameters are given in Marboeuf et al. (2012).

\section{Results}

Here, we have considered that the icy matrix made of clathrate and pure crystalline ice starts to devolatilize at temperatures higher than $\sim$47 K in the mixed model, which is the average clathrate formation temperature found in the PSN that matches the volatile content of 67P/C-G (Mousis et al. 2016). In the amorphous model, the icy matrix starts to devolatilize at temperatures higher than 130 K, corresponding to the crystallization temperature of amorphous water (Bar-Nun et al. 2007). The vastly different devolatilization temperatures imply that comets made from amorphous icy grains allow shorter formation delays than those made from crystalline ices and clathrates in the PSN to preserve their volatile budget. 

\begin{figure*}[h]
\begin{center}
\includegraphics[angle=0,width=16cm]{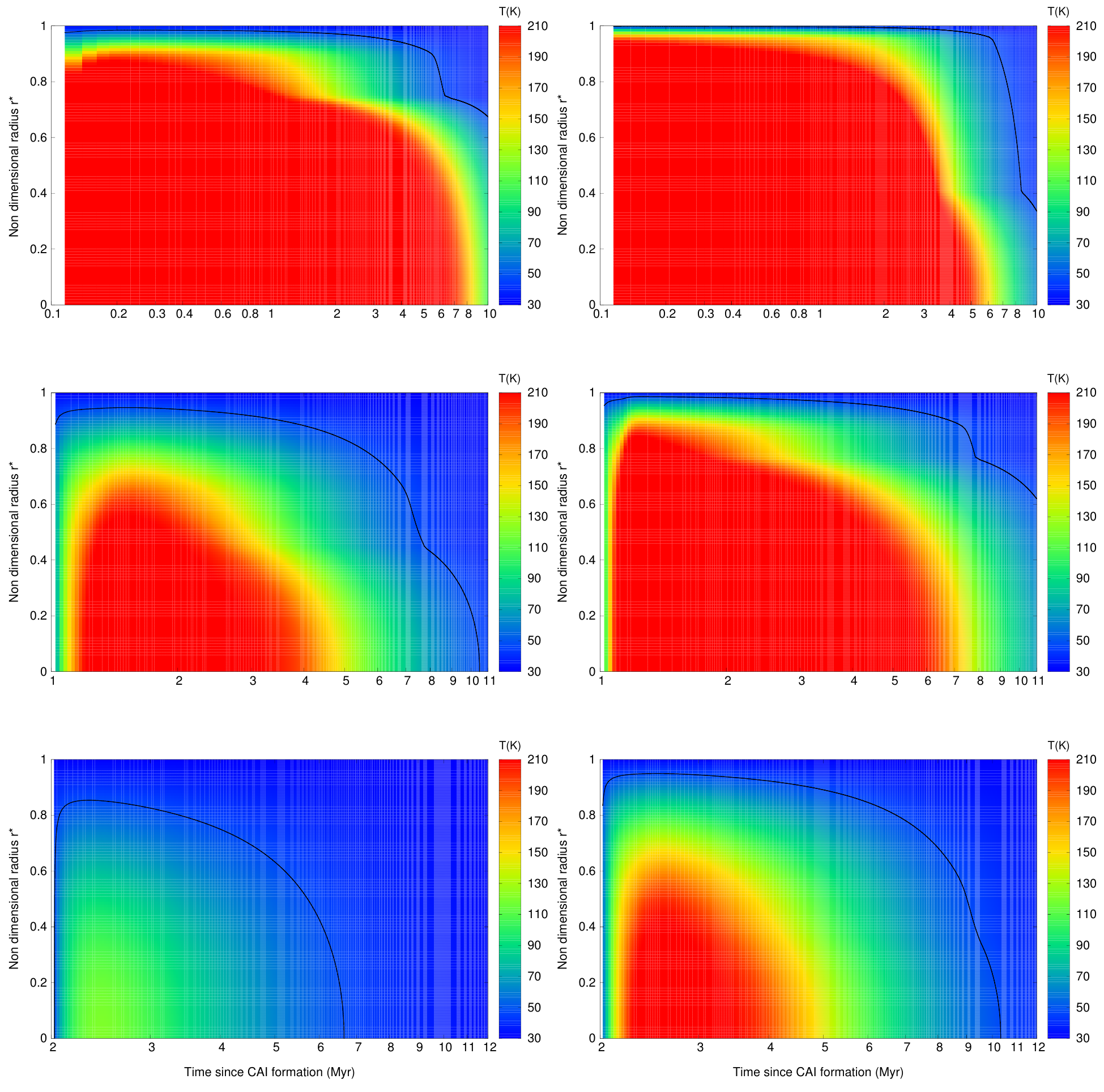}
\caption{From top to bottom: time evolution of the thermal profile within a body with a radius of 1.3 km after formation delays of 0, 1, and 2 Myr in the mixed model case. Left and right columns correspond to dust-to-ice ratios of 1 and 4 in the body, respectively. The black line corresponds to the 47 K isotherm representing the boundary between the stability and instability regions of the different ices, except water.}
\label{fig1}
\end{center}
\end{figure*}

\begin{figure*}[h]
\begin{center}
\includegraphics[angle=0,width=16cm]{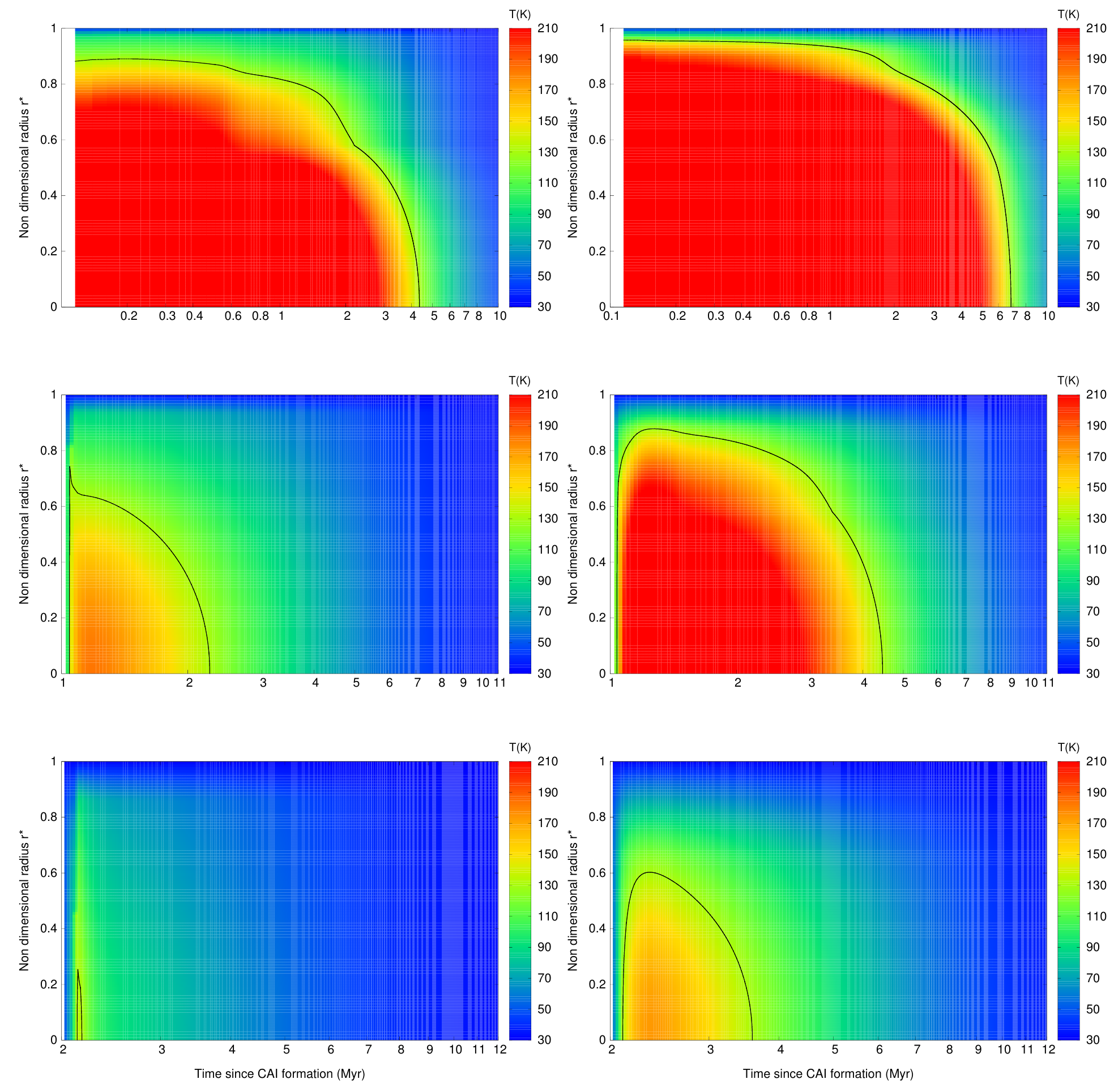}
\caption{Same as in Fig. \ref{fig2} but in the amorphous model case. The black line corresponds to the 130 K isotherm representing the boundary between the stability and instability regions of the amorphous ice.}
\label{fig2}
\end{center}
\end{figure*}

Figures \ref{fig1} and \ref{fig2} represent the time evolution of the thermal profile within a body with a radius of 1.3 km\footnote{Jorda et al. (2016) found that the individual lobes could be fitted by elipsoids whose principal axes of inertia are 4.10 $\times$ 3.52 $\times$ 1.63 km and 2.50 $\times$ 2.14 $\times$ 1.64 km. Our assumed radius of 1.3 km is an intermediary value between those ($\sim$1 and 1.5 km) inferred for each lobe, assuming they are fitted by spheres, and ignoring erosion.} in the cases of the mixed and amorphous models and for dust-to-ice~=~1~and~4, after formation delays $t_D$ of 0, 1, and 2 Myr. For the sake of clarity, we defined a non-dimensional radius $r^*$, corresponding to the value of $r/R$ with $R$ the total radius of the object. The two figures show that the body is warmer at a given epoch when the dust-to-ice ratio is set to 4, of course resulting from the larger mass fraction of the radioactive nuclides. In the case of the mixed model, the body becomes depleted in volatiles from the center up to $r^*$~=~0.98 (0.99), 0.95 (0.98), and 0.85 (0.95), assuming a dust--to--ice ratio of 1 (4), and for $t_D$ = 0, 1, and 2 Myr, respectively. These numbers can be translated in terms of volume fractions of preserved ices (outer shell of the body), which are 0.06 (0.03), 0.14 (0.06), and 0.39 (0.14) for the same values of dust--to--ice ratio and $t_D$. In the case of the amorphous model, the body becomes impoverished in volatiles from the center up to $r^*$~=~0.89 (0.95), 0.63 (0.88), and 0.24 (0.60), assuming a dust--to--ice ratio of 1 (4), and for $t_D$~=~0, 1, and 2 Myr, respectively. The corresponding volume fractions of preserved ices are 0.30 (0.14), 0.75 (0.32), and 0.99 (0.78) for the same values of dust--to--ice ratio and $t_D$.

Figure \ref{fig3} represents the extent of the devolatilized region as a function of the formation delay within bodies with radii of 1.3 and 35 km, respectively. It shows that the accretion of a typical lobe of 67P/C-G must start at least between $\sim$2.2 and 2.5 Myr after CAI formation for dust-to-ice ratios of 1 and 4, respectively, to fully preserve its volatile content in the case of the amorphous model. With values of $\sim$3.4 and 4.4 Myr for dust-to-ice ratios of 1 and 4, respectively, formation delays become longer in the case of the mixed model, as a result of its higher thermal inertia. The figure also shows that the formation delay of a Hale-Bopp sized body requires more time to fully preserve its volatile content. Here, to match this criterion, the body must start its accretion at least $\sim$5.6 and 5.9 Myr after CAI formation for dust-to-ice ratios of 1 and 4, respectively, in the case of the amorphous model. Meanwhile, the accretion must start at least $\sim$7.3 and 7.7 Myr after CAI formation for dust-to-ice ratios of $\sim$1 and 4, respectively, in the case of the mixed model. The figure illustrates the obvious point that a larger body retains more heat than a smaller one because its heat dissipation is proportional to $R^2$ while the total radiogenic heat produced increases as $R^3$.

\section{Influence of accretion time span}

Our thermal evolution model does not account for the accretion phase of the body. However, the time span taken by accretion remains an unconstrained parameter that can also affect the thermal history of planetesimals. Accretion may be very fast for silicate bodies, spanning no more than 100\,kyr in the case of the H parent body for instance (e.g. Monnereau et al. 2013a). However, the growth timescale is expected to increase with the orbital distance (Kokubo \& Ida 2000). 

While fast accretion yields bodies that experienced a uniform maximum temperature except within a superficial thermal boundary layer, slow accretion preserves a thicker mantle from heating by wrapping a smaller metamorphosed interior. We also investigated this effect through calculations with an energy conservation equation that takes into account accretion spread out over a long period (see Monnereau et al. 2013a for modeling details), but with the energy source term restricted to the radiogenic contribution $Q^{rad}$; i.e. without the complexity of phase changes for gas and ices. We found that the devolatilizated nucleus has almost the same radius for instantaneous accretion after a delay $t_D$ as for an accretional onset at CAI condensation and spanning a time $\tau=t_D$.  More precisely, Figure \ref{fig4} shows that bodies with a radius of 1.3\,km (35\,km) preserve 50\% of their volume from devolatilization  if $\tau > 2\textrm{ Myr} - t_D$ ($\tau > 7\textrm{ Myr} - t_D$), both relations being obtained for the mixed model with a dust-to-ice mass ratio of 4. As a consequence, the question of accretion delay of comets can also be posed in terms of accretion time span, since both processes and their interplay remain poorly understood.

\section{Discussion and conclusions}

Our computations support the conclusion that 67P/C-G's volatile content can either be explained via its agglomeration from building blocks originating from the PSN or from debris resulting from the disruption of a larger body having a Hale-Bopp size. Each composition case considered can be matched by these two scenarios via a particular set of plausible values for the formation delay $t_D$ and the extent of the devolatilized region. Our calculations show that, to fully retain their initial volatile budget, the two lobes of 67P/C-G must have accreted in the 2.2--4.4 Myr range after CAI formation, depending on the adopted type of ice and dust-to-ice ratio, and assuming they assembled from building blocks originating from the PSN. If 67/C-G's lobes assembled from chunks issued from the disruption of a parent body having a size similar to that of Hale-Bopp, their accretion time is delayed to 5.6--7.7 Myr after CAI formation, based on similar assumptions. Because accretion occurring during a given time span induces a similar extent of devolatilization within the body as instantaneous accretion happening at the end of the time span {\bf (assuming the negligible effect of phase transition in one case)}, the aforementioned numbers also correspond to the accretion time taken by the comet/parent body to reach its current/original size.

\begin{figure}[h]
\begin{center}
\resizebox{\hsize}{!}{\includegraphics[angle=0]{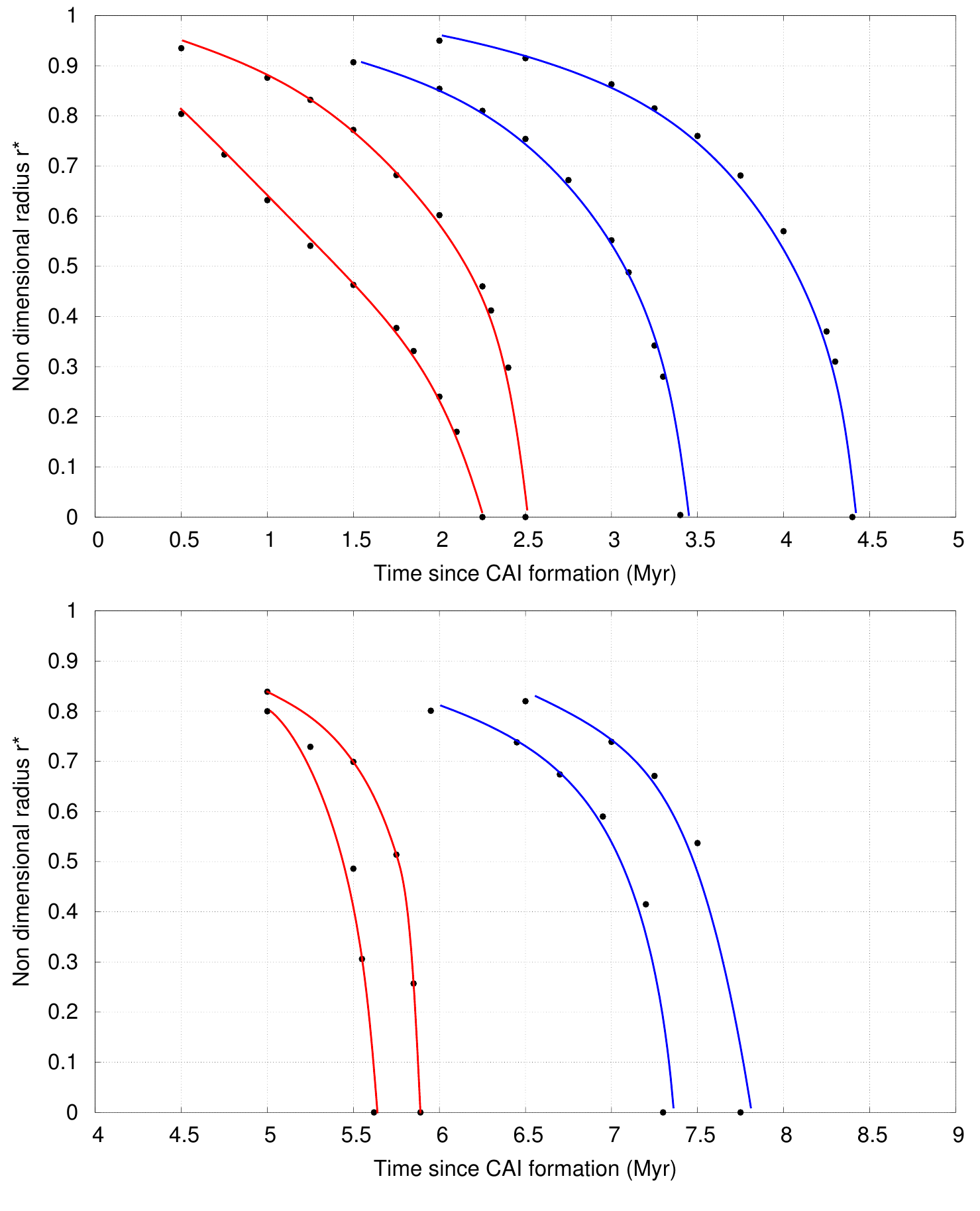}}
\caption{Extent of the devolatilized region (between 0 and $r^*$) within a body with a radius of 1.3 km (top panel) and within a Hale-Bopp sized body (bottom panel) as a function of its formation delay (D/I stands for dust--to--ice ratio). The red and blue curves correspond to the amorphous and mixed models, respectively. }
\label{fig3}
\end{center}
\end{figure}

\begin{figure}[h]
\begin{center}
\resizebox{\hsize}{!}{\includegraphics[angle=0]{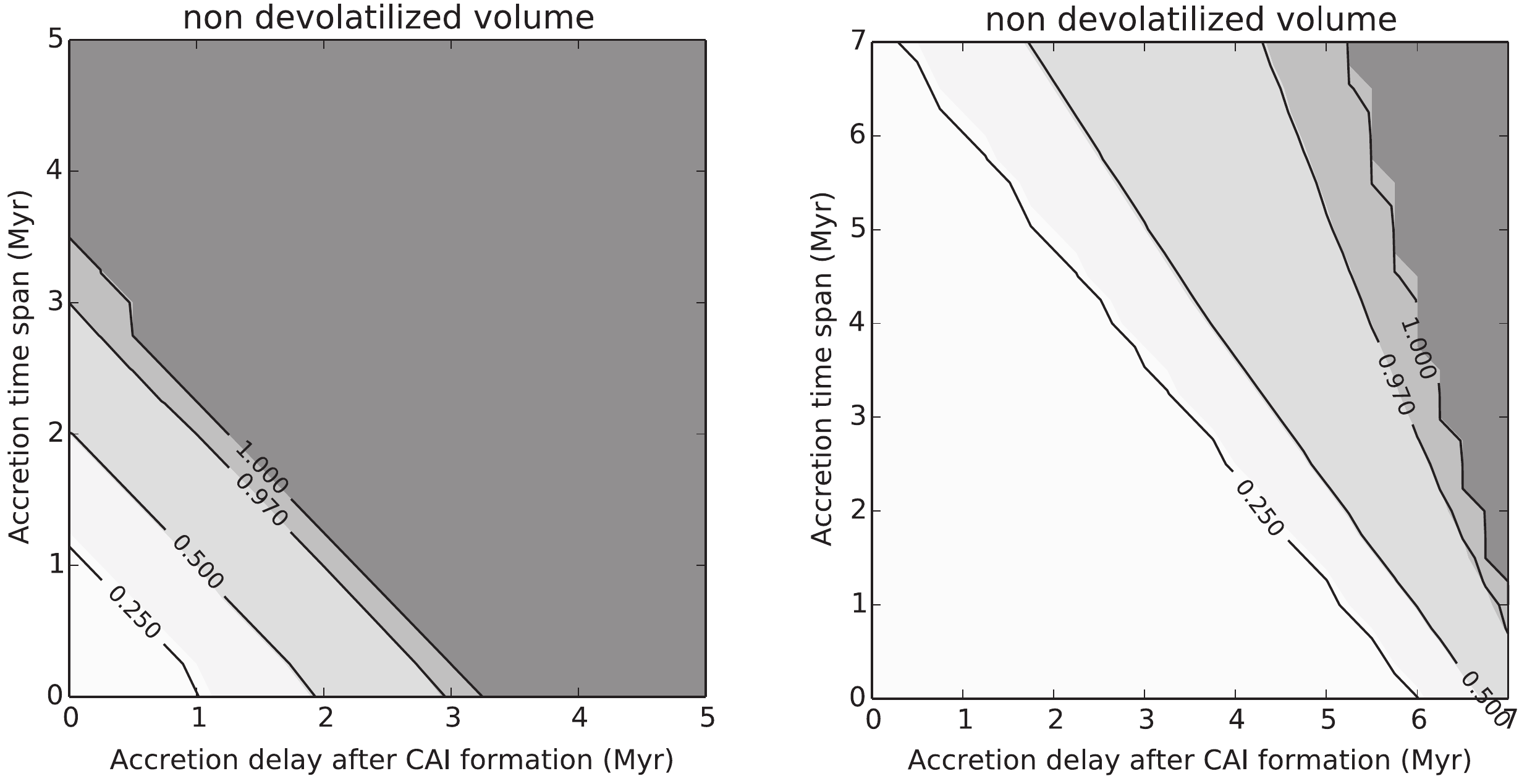}}
\caption{Non-devolatilized volumes within a body with a radius of 1.3 km (left panel) and within a Hale-Bopp sized body (right panel) represented as a function of accretion delay after CAI formation and accretion time span.}
\label{fig4}
\end{center}
\end{figure}

Shorter formation delays (or accretion time spans) can remain consistent with the {\it Rosetta} observations of ultravolatiles in 67P/C-G's environment, depending on the extent of the devolatilized region. In particular, thermo-physical models suggest that the comet could have lost a surface layer of up a few hundred meters thickness due to the accumulated activity during the course of its orbital evolution (Sierks et al. 2015; Keller et al. 2015; Rickman et al. 2015), implying that the molecules sampled by the ROSINA mass spectrometer aboard {\it Rosetta} come from $r^*$ values relatively close to 1. Assuming that only the deeper layers of 67P/C-G were devolatilized due to radiogenic heating, the comet could have formed over shorter timescales. For example, Fig. \ref{fig3} shows that, 67P/C-G's icy matrix remains pristine above a $r^*$ value of $\sim$0.8 if it agglomerated from primitive building blocks 0.5 Myr after the formation of CAIs, in the case of the amorphous model and with a dust-to-ice ratio of 1. At similar conditions, the mixed model requires that 67P/C-G agglomerated $\sim$2.3 Myr after the formation of CAIs in order to retain volatiles in the same top 20\% of $r^*$. For a Hale-Bopp class comet, the delays are much longer, $\sim$6.1 and 6.7 Myr in the cases of the amorphous and mixed models, respectively. If the dust-to-ice ratio ratio is much larger, the delays are increased even more. However, if 67P/C-G was made from chunks coming from the disruption of a larger parent body, these solids may have been well mixed after the collision and it would be impossible to know if the chunks come from the outer or inner layers of the parent body. Interestingly, it has been found that the deficiency of calcium measured by the ROSINA instrument in 67P/C-G, compared to CI and CV meteorites, could be explained by the presence of aqueous alteration (Wurz et al. 2015). If the observed Ca deficiency is indeed caused by aqueous alteration, it contradicts Davidsson et al. (2016)'s statement claiming that comets do not show any sign of aqueous alteration, contrary to larger bodies. It also implies that 67P/C-G could have been agglomerated from a mixture of chunks coming from different parts of the parent body.

To investigate the influence of the entrapment of volatiles in 67P/C-G on its formation delay, we have performed a series of simulations restricted to a body with a 1.3 km radius and a dust-to-ice ratio of 1. Two distinct icy phases have been investigated. In the first case, the icy phase is made of half crystalline water and half CO clathrate, and in the second case, the icy phase corresponds to a mixture of amorphous ice with adsorbed CO. In both cases, CO fulfills the full trapping capacity of the ice ($\sim$10 and 17\% of CO in amorphous ice and clathrate, respectively). In the first case, our simulations show that the progression of the devolatilization front behaves similarly to the equivalent case without CO within the body, the time difference being negligible. In the second case, we notice that the adsorption of CO in the amorphous matrix delays the progression of the devolatilization front by 0.3 Myr on average. This delay difference, due to the consideration of the CO sublimation latent heat during crystallization, does not impact our conclusions. Similar conclusions should be found in simulations of the thermal evolution of larger bodies.

The initial abundance of radioactive isotopes is crucial for evaluating the thermal evolution of comets and parent bodies. It is possible that $^{26}$Al and $^{60}$Fe have not been distributed homogeneously in the PSN (Krot et al. 2012), implying that the nuclides abundances considered in our study are not correctly appropriated. Finally, the main conclusion of our work is that the question of the origin of 67P/C-G's building blocks remains unanswered. A sample return mission toward another JFC will be necessary to provide an insight, based especially on the investigation of the hydration level of the refractory material and the extent of radiogenic heating that took place in the body.

\acknowledgements
O.M. acknowledges support from CNES. J.I.L acknowledges support from JWST. J.H.W. acknowledges support from a Rosetta subcontract from NASA JPL (1296001). This work has been partly carried out thanks to the support of the A*MIDEX project (n\textsuperscript{o} ANR-11-IDEX-0001-02) funded by the ``Investissements d'Avenir'' French Government program, managed by the French National Research Agency (ANR). We thank L. Jorda for helpful comments about our work.

\newpage

\begin{table}[h]
\begin{center}  
\caption{Nucleus modeling parameters}         
{\footnotesize
\begin{tabular}{lll}     
\hline
\hline      
Parameter 										& Value 					& Reference					\\
\hline
Radius (67P/C-G single lobe) (km)						& 1.3 					& Jorda et al. (2016)				\\	
Radius (Hale-Bopp) (km)								& 35 						& Weaver \& Lamy (1997)			\\	
Dust density $\rho_d$								& 3000					& Marboeuf et al. (2012)			\\
Dust-to-ice mass ratio 								& 4						& Rotundi et al. (2015)  			\\
Density (kg/m$^3$) for D/I = 4							& 510 					& Jorda~et~al. (2014)			\\
Density (kg/m$^3$) for D/I = 1							& 340 					& This work					\\
Porosity (\%)										& 76						& Brugger~et~al. (2016) 			\\
Thermal inertia (W~K$^{-1}$~m$^{-2}$~s$^{1/2}$) for D/I = 4	& 100  					& Leyrat~et~al. (2015)			\\
\hline
Al abundance ($\chi_{Al}$) (\%)							& 1.75 (CV meteorites)		& Wasson \& Kallemeyn (1988)		\\
Fe abundance ($\chi_{Fe}$) (\%)						& 23.5 (CV meteorites)		& Wasson \& Kallemeyn (1988)		\\
$^{26}$Al/$^{27}$Al									& 5 $\times$ 10$^{-5}$ 		& Castillo-Rogez et al. (2007)		\\
$^{60}$Fe/$^{56}$Fe									& 1 $\times$ 10$^{-8}$ 		& Tang \& Dauphas (2015)		\\
$\tau_{rad}$ of $^{26}$Al (Myr)							& 1.05					& Castillo-Rogez et al. (2007)		\\
$\tau_{rad}$ of $^{60}$Fe (Myr)							& 3.75					& Rugel et al. (2009)				\\
$H_{rad}$ of $^{26}$Al (J kg$^{-1}$)						& 1.16 $\times$ 10$^{13}$	& Castillo-Rogez et al. (2009)		\\
$H_{rad}$ of $^{60}$Fe (J kg$^{-1}$)						& 4.35 $\times$ 10$^{12}$	& Castillo-Rogez et al. (2009)		\\
\hline
\end{tabular}}
\end{center}
\label{table}
\end{table}

\newpage

\newpage

\newpage

\newpage

\end{document}